# Visualization the electrostatic potential energy map of graphene quantum dots

Setianto Setianto, Liu Kin Men, Camellia Panatarani, and I Made Joni

View Online   Export Citation

## ARTICLES YOU MAY BE INTERESTED IN

The impact of blending speed and duration on the characteristics of powdered milk product X in PT. XYZ
AIP Conference Proceedings **2219**, 070001 (2020); https://doi.org/10.1063/5.0003104

Moisture sorption isotherm and shelf life of pumpkin and arrowroot starch-based instant porridge
AIP Conference Proceedings **2219**, 070002 (2020); https://doi.org/10.1063/5.0003102

The potential of NaCl elicitation on improving antioxidant capacity and functional properties of sprouted pigeon pea (Cajanus cajan) flour
AIP Conference Proceedings **2219**, 070005 (2020); https://doi.org/10.1063/5.0003642

# Visualization the Electrostatic Potential Energy Map of Graphene Quantum Dots

Setianto Setianto[1, 2, a)], Liu Kin Men[2], Camellia Panatarani[2, 3] and I Made Joni[1, 2, 3]

[1]*Biotechnology Study Program, Graduate School, Universitas Padjadjaran,*
[2]*Department of Physics, Universitas Padjadjaran,*
[3]*Nanotechnology and Graphene Research Centre, Universitas Padjadjaran,*
*Jl. Raya Bandung-Sumedang KM 21, Jatinangor, West Java 45363, Indonesia.*

a)Corresponding author: setianto@phys.unpad.ac.id

**Abstract.** Graphene quantum dots (GQDs) represent single layers up to dozens of graphene layers smaller than 30 nm. GQDs are newish molecules that have aroused great interest in research because of their exceptional and manageable optical, electrical, chemical, and structural properties. In this work, we report electrostatic potential energy maps, or molecular electrostatic potential surfaces, illustrate the charge distributions of GQDs three-dimensionally. Knowledge of the charge distributions can be used to determine how GQDs interact with one another. To analyze the distribution of molecular charges accurately, a large number of electrostatic potential energy values must be calculated.The best way to transmit these data is to visualize them as in the electrostatic potential map. A ZINDO semi-empirical quantum chemistry method then imposes the calculated data onto an electron density model of the GQDs derived from the Schrödinger equation. To make the electrostatic potential energy data of GQDs easy to interpret, a color spectrum, with red as the lowest electrostatic potential energy value and blue as the highest, is employed to convey the varying intensities of the electrostatic potential energy values.The results of the four GQD system quantum models suggest that the energy of the ionization potential lies in a range of -7.20 eV to -5.31 eV and the electron affinity is -2.65 to -0.24 eV.

## INTRODUCTION

At present, carbon-based materials, especially graphene and its derivatives, such as graphene oxide (GO); reduced graphene oxide (rGO) and Graphene Quantum Dots (GQDs) have stimulated intensive research efforts for some interdisciplinary science, which covers various scientific disciplines, including chemistry, material science, physics, and nanotechnology [1-3]. In addition, graphene and its derivatives have revolutionized scientific developments in biomedicine, such as in drug delivery systems, bio-imaging and cancer therapy [4,5]. However, the graphene zero bandgap limits its practical application in optoelectronics and photonics [6]. Surface modification, doping and reduction of lateral dimensions of graphene to nano-bands and / or quantum dots (QDs) are considered as the main approach for handling band gap phenomena [7]. And also especially the enhanced Quantum Restriction (QCE) effect, the edge effect becomes clear when Graphene is converted to GQDs. This extraordinary property gives GQD new physical properties for a variety of applications above. GQD is a class of zero-dimensional nano graphite material with a lateral dimension of less than 100 nm and can be both single-layer and multilayer [1,8,9]. This GQD in terms of chemical inertness, ease of manufacture, resistance to photobleaching, low cytotoxicity and excellent biocompatibility compared to conventional semiconductor QD. Therefore, they are suitable for sensors, bio-imaging, optoelectronic devices, etc. In addition, carboxyl and hydroxyl group's clusters at the edges allow them to have good water solubility and functionalization with a variety of organic and inorganic compounds or biological class. However, concrete applications of nanomaterials in biology and medicine are distinguished examined for biocompatibility [10]. Although GQDs has been considered for a variety of biological applications, such as tissue engineering, biotechnology, drug delivery, gene delivery, imaging and therapy and related toxicity issues. Computational chemistry follows the way to the main options for presentation molecules and their properties in a three-dimensional perspective have become branches of this chemistry. For their effectiveness and efficiency in the calculations and properties and exploration, like these GQDs molecules will behave in a reaction and make it into the most comprehensive and complete research. Thus, this research has a central role to play modeling of the surface GQD molecules that optimized with modern biotechnology tools i.e.computer chemistry for visualization of the molecular electrostatic potential energy surface. It is important to understand the surface and conformation of a GQDs molecule that is mapped to the electrostatic potential, as it allows us to predict the reactivity of the molecule. Koopman's theorem states that the ionization potential of a molecule is equal to the energy of the highest occupied molecular orbital, while the EA is equal to the energy of the lowest unoccupied molecular orbital. The main assumption behind Koopman's theorem is that the quantum mechanical states (or molecular orbital) of the system are unmodified when adding or removing one electron[11].The goal of the conformational analysis is to determine the most stable atomic orientations which provide basic important information for the discovery and development of new materials which increase the biological activity of existing drugs already on the market.

## METHODOLOGY

The geometry optimization study was conducted with Austin Model One (AM1), which was a semi-empirical method using the ArgusLab 4.0.1 software. The minimum potential energy was calculated by the ZINDO/S method and the electrostatic potential mapped on the electron density surface of GQDs was generated. The ZINDO/S method was developed primarily for calculation excitation energies and ionization potential energy. The method cannot be used to optimize molecular geometries or to look for transition state structures. To do so, ZINDO/S calculations of a molecule one should first find the experimental molecular geometry (X-Ray or NMR analysis) or calculation with the standard Quantum Mechanics (QM) or Molecular Mechanics (MM) methods. Note that ZINDO/S like other semi-empirical methods suffers from occasional outliers and therefore provisional test calculations are required to complete the implementation of the method for systems of interest.The resulting data is used to image the surface of molecular orbitals and electrostatic potential mapped on the electron density [12,13]. The surface is used to visualize the ground state and the excited states such as orbitals, electron density and electrostatic potential (ESP). Hereby we present an electrostatic potential mapped on electron density surface and also calculated the ionization potential energy and electron affinity of GQDs system i.e.,$C_{24}H_{12}$ (coronene), $C_{30}H_{14}$ (dibenzo[bc,kl]coronene), $C_{18}H_{12}$ (triphenylene) and $C_{22}H_{12.}$

## RESULTS AND DISCUSSION

The electrostatic potential and the electrostatic potential mapped on the electron density surface of GQDs systems are shown in Figs. 1 and 2, respectively. The ionization potential energy and electron affinity of GQDs are presented in Tables 1 and 2, respectively. ArgusLab software is used to make and mapped the surface. This is the surface where one property is superimposed on the surface made by another property. The most preferred example of this is generally electrostatic potential (ESP) mapped on the surface of electron density. On ESP-mapped density surfaces, electron density surfaces offer surface shapes whereas ESP offers color.

The potential energy felt by the positive test charge at a particular destination in the area is known as the electrostatic potential. Negative ESP is often stability for test loads. Conversely, a positive ESP is often about relative instability for a positive test load. Thus, the molecular preference for nucleophilic or electrophilic attacks can be seen using ESP-mapped surface densities. This surface is useful for qualitative interpretations of chemical reactivity. In a different way to think about the density mapped by ESP, the surface shows that the electron density is the border for the largest (or at least) molecule relative to the nuclei. In the surface density of ESP-GQDs (Figure 2), the color shows ESP on the surface of electron density. The red color on the surface indicates the region with high electron density. Nucleophilic attacks are likely to occur in this region. Electrophilic attacks will occur in this region. The white part on the surface is a hydrogen group except for $C_{22}H_{12}$ system which has an unusual surface which will be further investigated intensively.

We also have calculated the ionization energy and electron affinity for the GQDs system selection and then compared it with available experimental data. In all of GQDs systems, we find that the calculation of ionization energy and electron affinity is not problematic, whereas the calculation of EA is much more difficult to compare with experimental data.The results of the four GQD system quantum models suggest that the energy of the ionization potential lies in a range of -7.20 eV to -5.31 eV and the electron affinity is -2.65 to -0.24 eV.

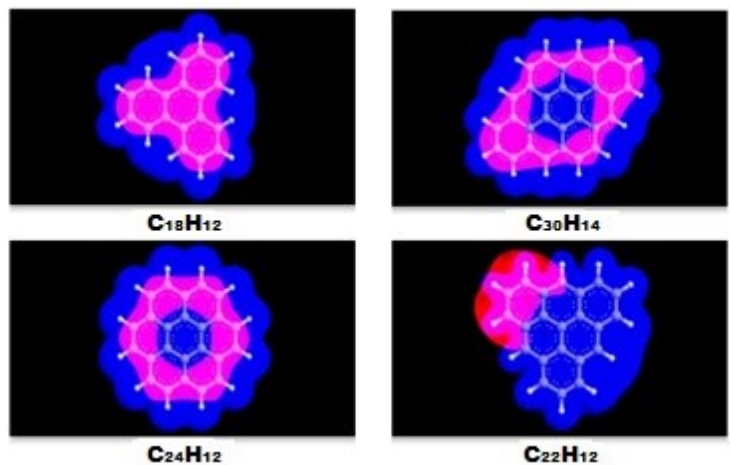
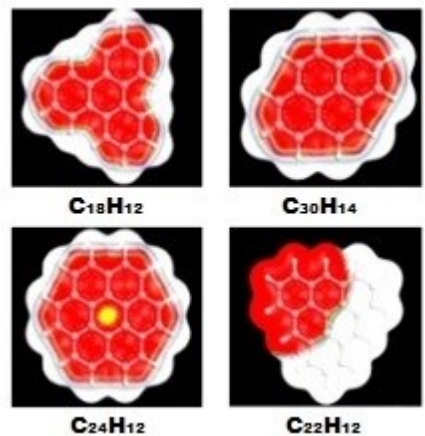

**FIGURE 1.** The electrostatic potential of GQDs systemi.e, $C_{24}H_{12}$ (coronene), $C_{30}H_{14}$(dibenzo[bc,kl]coronene), $C_{18}H_{12}$(triphenylene) and $C_{22}H_{12}$ withset the contour value is 0.05

**FIGURE 2.** The electrostatic potential mapped on electron density surface of GQDs system with set the contour value of density surface is 0.02

**TABLE 1.** The Ionization potential energy of GQD system

| GQDs system | $IE_{calc.}$ (eV) | $IE_{expt.}$ (eV) | Experimental Reference |
|---|---|---|---|
| $C_6H_6$ (benzene) | -9.14 | -9.24 | [14] |
| $C_{24}H_{12}$ (coronene) | -7.20 | -7.29 | [15] |
| $C_{30}H_{14}$ (dibenzo[bc,kl]coronene) | -6.32 | -6.42 | [16] |
| $C_{18}H_{12}$ (triphenylene) | -7.78 | -7.84 | [17] |
| $C_{22}H_{12}$ | -5.31 | | |

**TABLE 2.** The Electron affinity of GQD system

| GQDs system | $EA_{calc.}$ (eV) | $EA_{expt.}$ (eV) | Experimental Reference |
|---|---|---|---|
| $C_6H_6$ (benzene) | -0.71 | - | |
| $C_{24}H_{12}$ (coronene) | -0.73 | -0.47 | [18] |
| $C_{30}H_{14}$ (dibenzo[bc,kl]coronene) | -1.62 | - | |
| $C_{18}H_{12}$ (triphenylene) | -0.24 | -0.14 | [19] |
| $C_{22}H_{12}$ | -2.65 | - | |

# CONCLUSION

We have presented a convenient, simple, and computationally inexpensive procedure that allows one to determine the ionization energies, electron affinities, and visualize the electrostatic potential energy map of GQDs systems with accuracy comparable to experimental data.

# ACKNOWLEDGMENTS

This work is supported by Academic Leadership Grant (ALG), Contract Number 3498/UN6.D/LT/2019 and RDDU Project 2019 (3372/UN6.D/LT/2019).

# REFERENCES

1. S. Stankovich, D. A. Dikin, G. H. Dommett, K. M. Kohlhaas, E. J. Zimney, E. A. Stach, R. D. Piner, S. T. Nguyen and R. S. Ruoff, Nature **442**, 282–286 (2006).
2. A. C. Ferrari, F. Bonaccorso, V. Fal'Ko, *et al.*, Nanoscale **7**(11), 4598–4810 (2015).
3. A. K. Geim and K. S. Novoselov, Nat. Mater. **6**, 183–191 (2007).
4. M. Nurunnabi, K. Parvez, M. Nafiujjaman, V. Revuri, H. A. Khan, X. Feng and Y. Lee, RSC Adv. **5**, 42141–42161 (2015).
5. L. Lin, M. Rong, F. Luo, D. Chen, Y. Wang and X. Chen, Trends Anal. Chem. **54**, 83–102 (2014).
6. I. Meric, M.Y.Han, A.F. Young, B. Ozyilamz, P. Kim and K. Shepard, Nat. Nanotechnol. **3**(11), 654–659 (2008).
7. Y. Dong, H. Pang, H. B. Yang, C. Guo, J. Shao, Y. Chi, C. M. Li and T. Yu, Angew. Chem. Int. Ed.**52** (30), 7800–7804 (2013).
8. L. A. Ponomarenko, F. Schedin, M. I. Katsnelson, R. Yang, E. W. Hill, K. S. Novoselov and A. K. Geim, Sci. **320** (5874), 356–358 (2008).
9. V. Štengl, S. Bakardjieva, J. Henych, K. Lang, M. Kormunda, Carbon **63**, 537–546 (2013).
10. H. Dong, C. Dong, T. Ren, Y. Li, D. Shi, J. Biomed. Nanotechnol. **10**(9), 2086–2106 (2015).
11. Koopmans and Tjalling, Physica. **1**(1–6), 104–113 (1934).
12. J. Simons, P. Jorgensen, H. Taylor and J. Ozment, J. Phys. Chem. **87**, 2745–2753 (1983).
13. I. G. Csizmadia and R. D. Enriz, Theochem **543**, 319–361 (2001).
14. G. I. Nemeth, H. L. Selzle and E. W. Schlag, Chem. Phys. Lett. **215**(1–3), 151–155 (1993).
15. E. Clar, J. M. Robertson, R. Schloegl and W. Schmidt, J. Am. Chem. Soc. **103**(6), 1320–1328 (1981).
16. Clar and Schmidt, Tetrahedron **34**(21), 3219–3224 (1978).
17. Mautner (Meot-Ner), Pacific J. Math. **86**(1), 155–169(1980).
18. M. A. Duncan, A. M. Knight, Yuichi Negishi, S. Nagao, Y. Nakamura, A. Kato, A. Nakajima, K. Kaya, Chem. Phys. Lett. **309**(1–2), 49–54 (1999).
19. W. E. Wentworth and R. S. Becker, J. Am. Chem. Soc. **84**(22), 4263–4266 (1962).